\documentclass[twocolumn,showpacs,preprintnumbers,amsmath,amssymb]{revtex4}
\usepackage{amsmath,amssymb,graphics,epsfig,subfigure}
\usepackage{color}
\usepackage[colorlinks,
            linkcolor=blue,
            anchorcolor=blue,
            citecolor=green
            ]{hyperref}

\begin{document}
\renewcommand{\baselinestretch}{1.3}

\title{Constraints on generalized Eddington-inspired Born-Infeld branes}

\author{Zi-Chao Lin$^a$\footnote{linzch16@lzu.edu.cn},
        Ke Yang$^c$\footnote{keyang@swu.edu.cn},
        Yu-Peng Zhang$^a$\footnote{zhangyupeng14@lzu.edu.cn},
        Jian Wang$^a$\footnote{wangjian16@lzu.edu.cn},
        Yu-Xiao Liu$^a$$^b$\footnote{liuyx@lzu.edu.cn, corresponding author}}

\affiliation{$^{a}$Institute of Theoretical Physics $\&$ Research Center of Gravitation, Lanzhou University, Lanzhou 730000, China\\
$^{b}$Key Laboratory for Magnetism and Magnetic of the Ministry of Education, Lanzhou University, Lanzhou 730000, China\\
$^{c}$School of Physical Science and Technology, Southwest University, Chongqing 400715, China}

\begin{abstract}
The Palatini $f(|\hat{\Omega}|)$ gravity is a generalized theory of the Eddington-inspired Born-Infeld gravity, where $\Omega_{~N}^{K}\equiv\delta_{~N}^{K}+bg^{KL}R_{LN}(\Gamma)$ is an auxiliary tensor constructed with the spacetime metric $g$ and independent connection $\Gamma$. In this paper, we study $f(|\hat{\Omega}|)$ theory with $f(|\hat{\Omega}|)=|\hat{\Omega}|^{\frac{1}{2}+n}$ in the thick brane scenario and give some constraints on the brane model. We finally found an analytic solution of the thick brane generated by a single scalar field. The behavior of the negative energy density denotes the localization of the thick brane at the origin of the extra dimension. In our braneworld, the warp factor is divergent at the boundary of the extra dimension while the brane system is asymptotically anti$-$de Sitter. It is shown that the tensor perturbation of the brane is stable and the massless graviton is localized on the thick brane. Therefore, the effective Einstein-Hilbert action on the brane can be rebuilt in the low-energy approximation. According to the recent test of the gravitational inverse-square law, we give some constraints on the $f(|\hat{\Omega}|)$ brane.
\end{abstract}

%\keywords{Black holes, critical phenomena, phase diagram}


\maketitle

\section{Introduction}

As the most successful gravitational theory, general relativity (GR) is able to explain the gravitational phenomena in the scale ranging from submillimeter to Solar System scales~\cite{Will1}, for example, the deflection of light~\cite{Dyson1}, the precession of Mercury's perihelion, and so on. In particular, the recently detected gravitational waves also provide strong evidence to support GR~\cite{Abbott1,Abbott2,Abbott4,Abbott5,Abbott6,Abbott7,Abbott8,Abbott9,Abbott10}. Though it has earned observational successes, there are still some issues beyond the ability of GR. From the theoretical point of view, GR is faced with an unavoidable singularity that appears during the process by which a star collapses into a black hole~\cite{Penrose1,Penrose2,Hawking1,Senovilla1,Joshi1,Joshi2}. Moreover, the singularity problem also occurs in the early Universe~\cite{Hawking2}. For both of these singularities, all the in-falling particles will be unavoidably destroyed, while the physical quantities, such as the curvature invariants, of the spacetime will diverge~\cite{Saini1}. Therefore, this kind of singularity is associated with the breakdown of the geodesics and is related to the incompleteness of the spacetime itself~\cite{Ellis1,Tipler1}. These issues indicate that GR might be an effective theory and should be modified on the regime of high energy. With this consideration, many modified theories of gravity came out and provided with us many new approaches to solve the long-standing problems in GR.

Recently, based on the pure metric Born-Infeld theory of gravity~\cite{Deser and Gibbons} and some related works~\cite{Born4,Born1,Born2,Born3,Eddington,Vollick}, a remarkable gravity called Eddington-inspired Born-Infeld (EIBI) gravity was introduced by Banados and Ferreira~\cite{Banados and Ferreira}. The action of this gravity is written as follows
\begin{eqnarray}
\mathcal{L}=\sqrt{-|g_{\mu\nu}|}\Big(\sqrt{|\delta_{~\lambda}^{\rho}+b\,g^{\rho\alpha}R_{\alpha\lambda}(\Gamma)|}-{\lambda}\Big).
\end{eqnarray}
By including a Palatini formalism in the action, the connection is independent of the spacetime metric, and hence this theory can avoid the ghosts as well as the singular solutions, such as the big bang singularity~\cite{Banados and Ferreira} and the charged black hole singularity~\cite{Olmo1,Wei1,Delsate}, successfully.
Moreover, the leading order correction of the action was found to be able to reproduce the Einstein-Hilbert action when the curvature was much smaller than the parameter $b$ in the theory~\cite{Beltran2}.

It was later found that, by introducing a new definition, $\Omega_{~\mu}^{\nu}\equiv\delta_{~\mu}^{\nu}+bg^{\nu\alpha}R_{\alpha\mu}(\Gamma)$, in the original EIBI theory, the Lagrangian density of the gravity part could be expressed as~\cite{Odintsov}
\begin{eqnarray}
\mathcal{L}=\sqrt{-|g_{\mu\nu}|}\big(|\hat{\Omega}|^{\frac{1}{2}}-{\lambda}\big).\label{LD1}
\end{eqnarray}
The gravitational field equations are given by
\begin{eqnarray}
q_{\mu\nu} &=& g_{\mu\nu} + bR_{\mu\nu}(\Gamma), \label{EiBI-c}\\
|\hat{\Omega}|^{\frac{1}{2}}q^{\mu\nu} &=& \lambda g^{\mu\nu} -b\kappa^{2}T^{\mu\nu}, \label{EiBI-d}
\end{eqnarray}
where $|\hat{\Omega}|^{1/2}=\sqrt{-q}/\sqrt{-g}$. It is obvious that Eqs.~\eqref{EiBI-c} and~\eqref{EiBI-d} are the same as the corresponding field equations in the original EIBI gravity. We should notice that, from Eq.~\eqref{EiBI-c} and the definition of the auxiliary tensor $\hat{\Omega}$, one could also define this auxiliary tensor as a contraction of the spacetime metric and the auxiliary metric $q_{\mu\nu}$, $\Omega^{\nu}_{~\mu}=g^{\nu\alpha}q_{\alpha\mu}$. It means that, in both cases, this auxiliary tensor could be regarded as a connection between the spacetime metric and the auxiliary metric. This $\hat{\Omega}$ representation of EIBI gravity shows a clear structure of EIBI gravity and one could easily construct its functional extensions. Since, from this point of view, the original theory could be regarded as a particular case of this extended gravity, a natural question comes out, that is, whether the nonsingular solution of the early Universe in EIBI gravity actually results from the particular form of the action. This question could be explained as the robustness of the predictions of the early Universe in the original theory~\cite{Odintsov}. Therefore, by extending the square root structure in the Lagrangian density~\eqref{LD1} to a general case, i.e. $f(|\hat{\Omega}|)=|\hat{\Omega}|^{\frac{1}{2}+n}$, Odintsov {\em et al.} introduced a generalized theory called $f(|\hat{\Omega}|)$ theory~\cite{Odintsov}. Recent researches has shown that this functional extension is able to present nonsingular solutions and shares similar properties with the original EIBI theory~\cite{Beltran1,Odintsov,Makarenko1,Bambi1}.

Inspired by the recent interesting research papers on EIBI branes~\cite{KeYang,QiMing,KeYang2,Fu1,Bazeia1,Jana1,Prasetyo1}, we aim to give some constraints on the parameters in this particular functional extension of EIBI gravity from its thick brane scenario. It could be feasible since, from the point of view of high-dimensional theories, each low-dimensional gravity could be recovered by a mathematical technique called dimensional reduction. Therefore, any constraint on the high-dimensional theories is related to the constraints in the corresponding four-dimensional reduced gravity. On the other hand, in the brane scenario, all the particles included in the Standard Model are constrained on the brane and GR could be recovered in a low-energy approximation on the brane. In other words, the well-known Newtonian gravitational potential could be recovered with a relatively large separation between the particles, while the contributions from the massive gravitons (which are introduced by the extra dimensions) would be apparent as long as the separation is small enough. It means that the recent test of the gravitational inverse-square law aiming to prove the Newtonian potential at the length scale of submillimeter~\cite{Jun1} could give us a possibility to constrain the free parameters in $f(|\hat{\Omega}|)$ brane. Since the four-dimensional $f(|\hat{\Omega}|)$ theory is indeed the four-dimensional reduced gravity of $f(|\hat{\Omega}|)$ brane, the constraints on the parameters in this four-dimensional theory could be further obtained. Moreover, according to these constraints, our model would give a lower bound on the mass of graviton Kaluza-Klein (KK) modes and an upper bound on the size of the extra dimension.

In this case, it is crucial to confirm whether it is available to construct the braneworld scenario in $f(|\hat{\Omega}|)$ gravity. Indeed, it requires at least the following tips. First, the energy density of the brane should not diverge and its behavior should coincide with the location of the brane in some particular regions. Second, with the existence of particles around the brane, the metric perturbations are required to be stable to avoid the collapse of the brane. Last but not least, the graviton zero mode should localize on the brane since, in the low-energy approximation on the brane, the effective Newtonian potential should be recovered. Here, we note that, compared with the discussions on the stability problem of the tensor perturbation of EIBI branes in Refs.~\cite{KeYang,QiMing}, we prove its stability in a more generic case.

This paper is organized as follows. In Sec.~\ref{3}, we give a brief review of EIBI theory and introduce its generalized theory together with a short discussion. In Sec.~\ref{5}, we construct the thick braneworld model and give the analytic brane solution. In Sec.~\ref{4}, we investigate the stability problem of tensor perturbation in a more generic case. We also study the localization of the graviton zero mode on the brane and then derive the four-dimensional effective theory of $f(|\hat{\Omega}|)$ theory in the low-energy approximation. After that, we investigate the correction to the usual Newtonian gravitational potential from the massive gravitons and give some constraints on the parameters in the gravity. Finally, in Sec.~\ref{6}, we give a short conclusion.

%\section{EiBI theory}\label{2}
\section{Background equations of $f(|\hat{\Omega}|)$ theory}\label{3}

We start with the action of the $D$-dimensional EIBI theory~\cite{Banados and Ferreira}
\begin{eqnarray}\label{aEiBI1}
S&=&\frac{1}{\kappa^{2}b}\int d^{D}x\Big[\sqrt{-|g_{MN}+bR_{MN}(\Gamma)|}\nonumber\\
&~&-{\lambda}\sqrt{-|g_{MN}|}\Big]+S_{m}(g_{MN},\phi),
\end{eqnarray}
where the constant $\kappa$ satisfies $\kappa^{2}$\,=\,8$\pi G$ with $G$ the $D$-dimensional Newtonian gravitational constant, $b$ is a free parameter in this theory, $\lambda$ is a parameter related to the cosmological constant, $g_{MN}$ is the spacetime metric, $R_{MN}(\Gamma)$ is the symmetric part of the Ricci tensor formed from the independent connection $\Gamma_{~MN}^{P}$, and $S_{m}(g_{MN},\phi)$ is the action of the matter fields coupled only to the metric $g_{MN}$. Note that, here and after, for the sake of simplicity, we use the notation $\Gamma$ to represent $\Gamma_{~MN}^{P}$. Being a Palatini formalism, the connection is dynamically independent of the spacetime metric, which leads to a variety of interesting geometric phenomena. The gravitational field equations have already been given in Eqs.~\eqref{EiBI-c} and~\eqref{EiBI-d}, where $q_{MN}$ is the auxiliary metric compatible with the independent connection $\Gamma_{~MN}^{P}=\frac{1}{2}q^{PK}\big(\partial_{M}q_{NK}+\partial_{N}q_{MK}-\partial_{K}q_{MN}\big)$. Note that $q^{MK}q_{ML}=\delta_{~L}^{K}$ with $q^{MN}$ the inverse of $q_{MN}$. In EIBI theory, there is no high-order term of the spacetime metric in the field equations, and hence it can avoid ghost instabilities~\cite{Beltran2}. The parameter $\lambda$ is related to the effective cosmological constant, $\Lambda_{\text{eff}}=\frac{\lambda-1}{2b}$. So, by setting $\lambda =1$, one can obtain an asymptotically Minkowski spacetime. Moreover, the theory will recover the Eddington gravity in the high curvature regime (called the ``Eddington regime,'' i.e., $R_{MN}\gg 1/b$) while recovering GR in the low curvature regime.

In Ref.~\cite{Odintsov}, the authors generalized EIBI theory to $f(|\hat{\Omega}|)$ theory with the action given by
\begin{equation}
S=\frac{1}{\kappa^{2}b}\int{d^{D}x\sqrt{-g}\Big[f(|\hat{\Omega}|)-{\lambda}\Big]}+S_{m}(g_{MN},\phi). \label{ea}
\end{equation}
It is clear that, with a specific choice of $f(|\hat{\Omega}|)=|\hat{\Omega}|^{1/2}$, this theory will recover EIBI theory at the action level. It should be noted that $f(|\hat{\Omega}|)$ theory can recover GR in the low-energy regime ($|bR_{MN}{\ll}1|$), while it will deviate from GR when the curvature is much larger than $1/b$. It is interesting to assume that
\begin{equation}
f(|\hat{\Omega}|)=|\hat{\Omega}|^{\frac{1}{2}+n},
\end{equation}
where the parameter $n$ vanishes for the case of EIBI theory. This assumption was first introduced in Ref.~\cite{Odintsov} with a slight modification of the index. The application of $f(|\hat{\Omega}|)$ theory in the early-time cosmology was discussed~\cite{Odintsov}. It was shown that this theory could always avoid big bang singularities without any fine-tuning. Two types of nonsingular solutions found in EIBI theory were also presented in $f(|\hat{\Omega}|)$ theory~\cite{Makarenko1}. In the black hole, by replacing the pointlike singularity with a wormhole structure, one can construct a modified inner structure of the black hole compared to the one in EIBI theory and hence obtain a nonsingular spacetime~\cite{Bambi1}. In this paper, we will consider the following matter action of a scalar field
\begin{equation}
S_{m}=\int{d^{D}x\sqrt{-g}\Big[-\frac{1}{2}g^{MN}\partial_{M}\phi\partial_{N}\phi-V(\phi)\Big]}.
\end{equation}
In order to compare the auxiliary metric in EIBI theory with the one in $f(|\hat{\Omega}|)$ theory, we define an auxiliary tensor
\begin{equation}
p_{MN}=g_{MN}+bR_{MN}(\Gamma).
\end{equation}
To obtain the field equations, we vary the full action:
\begin{eqnarray}
\delta S&=&\frac{1}{\kappa^{2}b}\int d^{D}x\Big[\frac{1}{2}\sqrt{-g}g^{MN}(f-\lambda)\delta g_{MN}\nonumber\\
&~& +\sqrt{-g}f_{|\hat{\Omega}|}|\hat{\Omega}|(\Omega^{-1})_{~K}^{P}\delta \Omega_{~P}^{K}\Big]+\delta S_{m},
\end{eqnarray}
where the variation of the matter action can be written as
\begin{eqnarray}
\delta S_{m}&=&\int d^{D}x
     \bigg\{\Big[\partial_{K}\Big(\sqrt{-g} \partial^{K}\phi\Big)
                 -\sqrt{-g} V_{\phi}
            \Big]\delta\phi\nonumber\\
            &~&+\sqrt{-g}
        \Big(\frac{1}{2} \partial^{M}\phi\partial^{N}\phi
             -\frac{1}{4}g^{MN}\partial^{K}\phi\partial_{K}\phi\nonumber\\
             &~&-\frac{1}{2}g^{MN}V
         \Big)
        \delta g_{MN}
     \bigg\} ,
\end{eqnarray}
with $V_{\phi}\equiv dV(\phi)/ d\phi$.
With the following relations
\begin{eqnarray}
 p_{MN}&=&g_{MK}\Omega_{~N}^{K},  \nonumber\\
  \delta R_{MN}(\Gamma)&=&\bigtriangledown_{K}^{(\Gamma)}(\delta\Gamma_{~MN}^{K})-\bigtriangledown_{N}^{(\Gamma)}(\delta\Gamma_{~KM}^{K}),\nonumber\\
 \delta\Omega_{~P}^{K}&=&\delta(p_{PL}g^{LK})\nonumber\\
 &=&g^{LK}\delta g_{PL}+bg^{LK}\delta R_{PL}\nonumber\\
 &~&-\Omega_{~P}^{M}g^{KN}\delta g_{MN},
\end{eqnarray}
the variation can be expressed as
\begin{eqnarray}
\delta S&=&\delta S_{m}+\frac{1}{\kappa^{2}b}\int d^{D}x\bigg\{\sqrt{-g}\Big[\frac{1}{2}g^{MN}\big(f-\lambda\big)\nonumber\\
&~&+|\hat{\Omega}|f_{|\hat{\Omega}|}p^{MN}-|\hat{\Omega}|f_{|\hat{\Omega}|}g^{MN}\Big]\delta g_{MN}\Big.\nonumber\\
\Big.&~&-b\bigtriangledown_{K}^{(\Gamma)}\Big(\sqrt{-g}|\hat{\Omega}|f_{|\hat{\Omega}|}p^{MN}\Big)\delta\Gamma_{~MN}^{K}\nonumber\\
&~&+b\bigtriangledown_{N}^{(\Gamma)}\Big(\sqrt{-g}|\hat{\Omega}|f_{|\hat{\Omega}|}p^{MN}\Big)\delta\Gamma_{~KM}^{K}\bigg\},
\end{eqnarray}
where the covariant derivative, $\bigtriangledown^{(\Gamma)}$, is compatible with the independent connection. By varying the full action with respect to the metric field, connection field, and scalar field, respectively, the equations of motion can be obtained as follows
\begin{eqnarray}
  0&=& \bigtriangledown_{K}^{(\Gamma)}
      \big(\sqrt{-p}|\hat{\Omega}|^{\frac{1}{2}}f_{|\hat{\Omega}|}p^{MN}\big)
       ,\label{am1} \\
  -b\kappa^{2}T^{MN}  &=& \big(f-\lambda\big)g^{MN}
     +2|\hat{\Omega}|f_{|\hat{\Omega}|}p^{MN}\nonumber\\
     &~&-2|\hat{\Omega}|f_{|\hat{\Omega}|}g^{MN}
     ,\label{Elf} \\
 V_{\phi}  &=& \frac{1}{\sqrt{-g}}\partial_{M}\big(\sqrt{-g}\partial^{M}\phi\big)\label{seom1}
                                ,
\end{eqnarray}
where the energy-momentum tensor is defined as
\begin{eqnarray}
T^{MN}&\equiv&\frac{2}{\sqrt{-g}}\frac{\delta S_{m}}{\delta g_{MN}}\nonumber\\
&=&\partial^{M}\phi\partial^{N}\phi-g^{MN}\Big(\frac{1}{2}\partial_{K}\phi\partial^{K}\phi+V\Big).
\end{eqnarray}
Equation~\eqref{Elf} is equivalent to
\begin{eqnarray}
-b\kappa^{2}T_{MN}&=&\big(f-\lambda\big)g_{MN}-2|\hat{\Omega}|f_{|\hat{\Omega}|}g_{MN}\nonumber\\
&~&+2|\hat{\Omega}|f_{|\hat{\Omega}|}g_{MK}g_{NL}p^{KL}.
\end{eqnarray}
Note that $p_{MN}\neq g_{MK}g_{NL}p^{KL}$. By defining another auxiliary tensor
\begin{equation}\label{am2}
q_{MN}\equiv\big(2|\hat{\Omega}|^{\frac{1}{2}}f_{|\hat{\Omega}|}\big)^{\frac{2}{D-2}}p_{MN},
\end{equation}
which leads to
\begin{eqnarray}
p^{MN} &=& \big(2|\hat{\Omega}|^{\frac{1}{2}}f_{|\hat{\Omega}|}\big)^{\frac{2}{D-2}}q^{MN},\\
|p_{MN}|&=&\big(2|\hat{\Omega}|^{\frac{1}{2}}f_{|\hat{\Omega}|}\big)^{\frac{-2D}{D-2}}|q_{MN}|,
\end{eqnarray}
one finds that Eq.~\eqref{am1} can be transformed as
\begin{equation}\label{am3}
\nabla_{K}^{(\Gamma)}\big(\sqrt{-q}q^{MN}\big)=0,
\end{equation}
or equivalently,
\begin{equation}\label{1111}
\nabla_{K}^{(\Gamma)} q^{MN} =0.
\end{equation}
Eq.~\eqref{1111} implies that the auxiliary tensor $q_{MN}$ is compatible with the independent connection $\Gamma$ referred before. Therefore, we will call the tensor $q_{MN}$ the auxiliary metric, and regard the definition $p_{MN}\equiv g_{MN}+bR_{MN}(\Gamma)$ as an equation of motion. For the case of EIBI theory, $f(|\hat\Omega|)=|\hat\Omega|^{1/2}$, Eqs.~\eqref{am2} and~\eqref{am3} will reduce to $q_{MN}\equiv p_{MN}$ and $\nabla_{K}^{(\Gamma)}\big(\sqrt{-p}p^{MN}\big)=0$, respectively. Therefore, the two tensors are the same in EIBI theory. It is worth noting that the auxiliary metric in $f(|\hat{\Omega}|)$ theory is usually different from that in EIBI theory, while they are related by Eq.~\eqref{am2}. Moreover, there exist some relations between the spacetime metric and the auxiliary one. The definition of the auxiliary metric
\begin{eqnarray}
q_{MN}\equiv\big(2|\hat{\Omega}|^{\frac{1}{2}}f_{|\hat{\Omega}|}\big)^{\frac{2}{D-2}} (g_{MN}+bR_{MN}(\Gamma)),
\end{eqnarray}
denotes the geometric relationship between them while Eq.~\eqref{Elf} implies that the two metrics are also connected by the matter field.

In this paper, we mainly focus on the five-dimensional $f(|\hat\Omega|)$ theory ($D=5$). To preserve the four-dimensional Poincar\'{e} invariance on the brane, we assume the metrics given as follows
\begin{eqnarray}
ds^{2}&=&g_{MN}dx^{M}dx^{N}=a^{2}(y)\eta_{\mu\nu}dx^{\mu}dx^{\nu}+dy^{2},\label{le1}\\
d\tilde{s}^{2}&=&q_{MN}dx^{M}dx^{N}\nonumber\\
&=&u(y)\eta_{\mu\nu}dx^{\mu}dx^{\nu}+v(y)dy^{2}.\label{le2}
\end{eqnarray}
Note that, in the Palatini formalism, the spacetime metric and the introduced auxiliary metric have different physical meanings and play different roles in defining the geometry of the spacetime. Therefore, one could define the above two kinds of line elements, i.e., Eqs.~\eqref{le1} and~\eqref{le2}, where the previous one could be seen by the matter field and the latter could be seen by the gravitational waves~\cite{Beltran2,Koivisto1,Sotiriou1,Beltran1,Jana2}. To clarify it, we first come to the matter field. It is obvious that, from the action~\eqref{ea}, the matter field only couples to the spacetime metric minimally. Therefore, the equivalent principle could be preserved well on the geometry defined by the spacetime metric and the free-falling particles should move along the geodesics defined by the Levi-Civit\`{a} connection compatible with the spacetime metric. Indeed, the action~\eqref{aEiBI1} was shown to have an equivalent form as~\cite{Beltran2}
\begin{eqnarray}\label{ea2}
\tilde{S}=\frac{1}{\kappa^{2}}\int{d^{D}x}\sqrt{-q}\,q^{MN}R_{MN}(\Gamma)+\tilde{S}_{m}(q_{MN},\phi),
\end{eqnarray}
where the matter field is nonminimally coupled with the auxiliary metric. Further, we could conclude that the similar equivalent form of the action~\eqref{ea} keeps the same properties. In this case, the equivalent principle is hard to preserve on the geometry defined by the auxiliary metric and the free-falling particles will instead feel a ``force'' from the effective energy-momentum tensor constructed by the auxiliary metric. For the gravitational waves, since the Einstein-Hilbert equation of the auxiliary metric could be obtained from the action~\eqref{ea2}, one may prefer to describe the propagations of them in the $q$ geometry~\cite{Beltran1,Jana2}.

Now, for the sake of simplicity, we replace the equations of motion~\eqref{am1} and~\eqref{Elf} as follows:
\begin{eqnarray}
g_{MN}+bR_{MN}(\Gamma)&=&\big(2|\hat{\Omega}|^{\frac{1}{2}}f_{|\hat{\Omega}|}\big)^{-\frac{2}{3}}q_{MN},\label{am4}\\
-b\kappa^{2}T^{MN}&=&\big(f-\lambda\big)g^{MN}-2|\hat{\Omega}|f_{|\hat{\Omega}|}g^{MN}\nonumber\\
&~&+2^{\frac{5}{3}}|\hat{\Omega}|^{\frac{4}{3}}f_{|\hat{\Omega}|}^{\frac{5}{3}}q^{MN}.\label{Elf2}
\end{eqnarray}

\section{Thick brane solutions}\label{5}

In this section, we mainly focus on the construction of thick branes in $f(|\hat{\Omega}|)$ theory. We will first give the background solutions with a short discussion and further analyze whether they are thick brane solutions. With the metric ansatz~\eqref{le1} and~\eqref{le2}, the explicit forms of Eqs.~\eqref{am4},~\eqref{Elf2}, and~\eqref{seom1} are shown as follows
\begin{eqnarray}
\frac{au^{\frac{1}{2}}}{2^{\frac{1}{4}}f^{\frac{1}{4}}_{|\hat{\Omega}|}v^{\frac{1}{8}}}&=&a^{2}+b\frac{uu'v'-2v(u'^{2}+uu'')}{4uv^{2}},\label{am5}\\
\frac{av^{\frac{7}{8}}}{2^{\frac{1}{4}}f^{\frac{1}{4}}_{|\hat{\Omega}|}u^{\frac{1}{2}}}&=&1+b\frac{u'^{2}}{u^{2}}+b\frac{u'v'-2u''v}{uv},\label{am6}\\
\frac{2^{\frac{3}{4}}u^{\frac{3}{2}}v^{\frac{3}{8}}}{af^{\frac{1}{4}}_{|\hat{\Omega}|}}&=&a^{2}\big[2(f-\lambda)-b\kappa^{2}(2V+\phi'^{2})\big]\nonumber\\
&~&+2uv^{\frac{1}{2}},\label{Elf3}\\
\frac{2^{\frac{3}{4}}u^{\frac{3}{2}}v^{\frac{3}{8}}}{af^{\frac{1}{4}}_{|\hat{\Omega}|}}&=&a^{2}\big[2(f-\lambda)-b\kappa^{2}(2V-\phi'^{2})\big]\nonumber\\
&~&+2\frac{u^{2}}{a^{2}v^{\frac{1}{2}}},\label{Elf4}\\
V_{\phi}&=&\frac{4a'\phi'}{a}+\phi'',\label{conservationlaw1}
\end{eqnarray}
where the primes denote the derivative with respect to the extra-dimensional coordinate $y$. It can be seen that there are six functions ($a$, $u$, $v$, $f$, $\phi$, and $V$) but with five equations. However, the law of local energy-momentum conservation breaks the independence among the last three equations. So one can usually use two of them to express the other one and finally one has four independent equations with six functions. Now, we have two superfluous variables, and therefore we can introduce some relations among these six functions, or even some assumptions of the forms of them. In this paper, we regard the expression $f(|\hat{\Omega}|)=|\hat{\Omega}|^{\frac{1}{2}+n}$ as an assumption. Further, to coincide with the thick brane model, we assume that the warp factor has the typical form of $a(y)=\text{sech}^{m}(ky)$. After a tedious calculation, the background solution is finally obtained:
\begin{eqnarray}
a(y)&=&\text{sech}^{\frac{1}{4}-\frac{n}{2}}(ky),\label{a1}\\
u(y)&=&K\,\text{sech}^{\frac{1}{2}+\frac{n}{3}}(ky),\label{u1}\\
v(y)&=&\frac{2nK}{3+2n}\text{sech}^{2+\frac{4n}{3}}(ky),\label{v1}\\
\phi(y)&=& \frac{(2n+3)^{\frac{3}{4}}}{2^{\frac{5}{4}}n^{\frac{1}{4}}\kappa}kK^{\frac{3}{4}}\nonumber\\
&~&           \times\int \sqrt{1+\frac{2n}{3}\text{tanh}^{2}(ky)}
                \text{sech}^{n-\frac{1}{2}}(ky) dy,\label{phi1}\nonumber\\
                &~&\\
V(y)&=&\frac{\sqrt{2} K^{\frac{3}{2}} ~  \text{sech}^{2n-1}(ky) }
              {4b\kappa^{2}\sqrt{n}\sqrt{2n+3}(2n+1)}\big[4n(n+2)\nonumber\\
&~&          +3+2n(3-2n)\text{sech}^{2}(ky)\big]-\frac{\lambda}{b\kappa^{2}}
         ,\label{V1}
\end{eqnarray}
where
\begin{eqnarray}
k&=&\frac{2}{2n+3}\sqrt{-\frac{3}{b}},\\
K&=& \frac{(2n-1)^{\frac{10n}{3}+1}  (2n+1)^{\frac{2}{3}}  }
          {2^{\frac{8n}{3}+1}   n^{\frac{8n}{3}+1} (2n+3)^{\frac{2n}{3}} },\label{K1}
\end{eqnarray}
and $b$ is fixed as a negative to guarantee that the parameter $k$ is real. One can find that, for the given background solution, there are some constraints on $n$, i.e., $n\neq0$, $\pm1/2$, $-3/2$. It seems that the constraints on $n$ prevent the $f(|\hat{\Omega}|)$ theory considered here from recovering EIBI theory. However, the constraints are actually arisen from the particular assumptions we have chosen to eliminate the above two superfluous variables, or, on the other hand, their existences depend on our procedure to solve the field equations. So, they may vanish if one uses other ways to solve the equations of motion. In addition, if $n=0$, or equivalently $f=|\hat{\Omega}|^{\frac{1}{2}}$, using the assumptions given by Liu {\em et al.}~\cite{KeYang,QiMing}, there should exist a family of thick brane solutions. Although we finally obtain an unshown analytical solution of the scalar field by integrating Eq.~\eqref{phi1}, we find that there exists the Appell series $F_{1}$ in the solution of (\ref{phi1}), from which we can conclude that the solution is finite only for the case of $n>1/2$.  However, because of the existence of the hypergeometric function, it is hard to obtain the analytic solution of the scalar potential $V(\phi)$. One can choose some values, such as $5/2$, $9/2$, and $13/2$, of the parameter $n$ to obtain the simple analytical expressions of the scalar field, which are given as follows:

\begin{eqnarray}
\phi(y)&=&A\sqrt{\frac{5\,\text{tanh}^{2}(ky)+3}{4\,\text{cosh}(2ky)-1}}\nonumber\\
&~&\times\Bigg[5\sqrt{4\,\text{cosh}(2ky)-1}~\text{tanh}(ky)\Big.\nonumber\\
\Big.&~&+3\sqrt{5}~\text{arctanh}\Bigg(\sqrt{\frac{5\,\text{sinh}^{2}(ky)}{4\,\text{cosh}(2ky)-1}}\,\Bigg)
       \,\text{cosh}(ky)\Bigg],\nonumber\\
       &~&(n=5/2)\\
\phi(y)&=&B\,\text{cosh}(ky)\sqrt{\frac{30\,\text{tanh}^{2}(ky)+10}{2\,\text{cosh}(2ky)-1}}\nonumber\\
&~&\times          \Bigg[13\,\text{arctanh}\Bigg(\sqrt{\frac{3\,\text{sinh}^{2}(ky)}{2\,\text{cosh}(2ky)-1}}\,\Bigg)\nonumber\\
       &~&+\sqrt{6\,\text{cosh}(2ky)-3}\big(6\,\text{sech}^{2}(ky)+5\big)\nonumber\\
       &~&\times\text{sech}(ky)\,\text{tanh}(ky)\Bigg],\nonumber\\
&~&(n=9/2) \\
\phi(y)&=& C_{1}\,\text{sech}^{5}(ky)
      \sqrt{\frac{13\,\text{tanh}^{2}(ky)+3}{8\,\text{cosh}(2ky)-5}}\nonumber\\
      &~&\times\Bigg[ C_{2}\,\text{arctanh}
       \Bigg(\sqrt{\frac{13\,\text{sinh}^{2}(ky)}{8\,\text{cosh}(2ky)-5}}\,\Bigg)\,\text{cosh}^{6}(ky)\Big.\nonumber\\
\Big.&~&+13\sqrt{8\,\text{cosh}(2ky)-5} \sum_{i=1}^{3} D_i\,\text{sinh}\big((2i-1)ky\big) \Bigg],\nonumber\\ &~&(n=13/2)
\end{eqnarray}
where
\begin{eqnarray}
A&=&\frac{2048}{78125\,\kappa},~B=\frac{536870912}{94143178827\,\kappa},\nonumber\\
C_{1}&=&\frac{470184984576\sqrt{42}}{8650415919381337933\,\kappa},\nonumber\\
C_{2}&=&218448\sqrt{13}, \nonumber \\
D_{1}&=&28598,~D_{2}=7901,~D_{3}=935.
\end{eqnarray}
Note that, on the boundary of the extra dimension, the scalar field approaches to a constant, i.e., $\phi|_{y\rightarrow\pm\infty}=\pm v_{0}=h(n)(2n+3)^{3/4}K^{3/4}/(2^{5/4} \kappa n^{1/4})$, where $h(n)$ is a function of the parameter $n$ only. For the case of $n=5/2$, $n=9/2$, and $n=13/2$, $h(n)$ equal to $1.23205$, $0.828\,521$, and $0.664\,938$, respectively.

\begin{figure}[!htb]
\center{
\subfigure[$a(y)$]{\includegraphics[width=4cm]{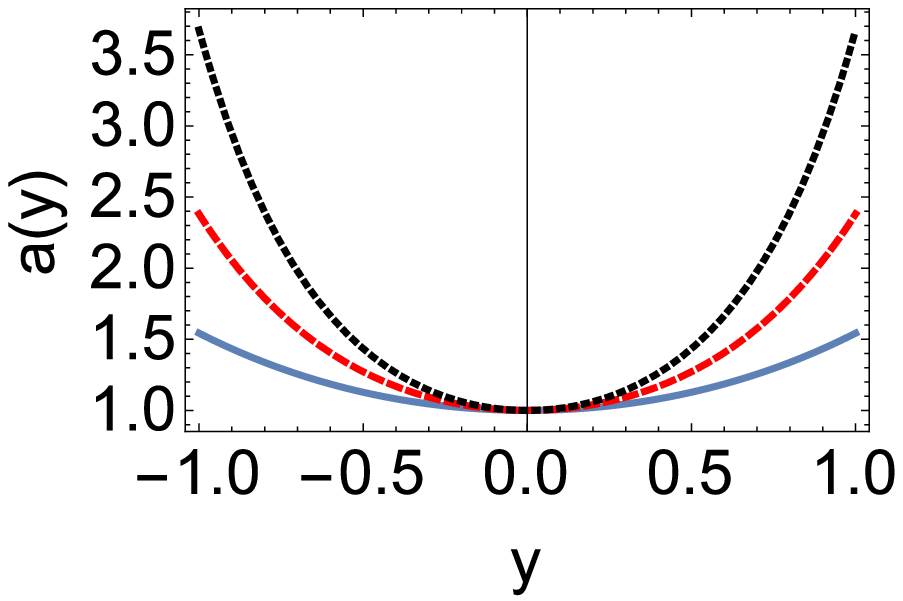}}
\subfigure[$\phi(y)$]{\includegraphics[width=4cm]{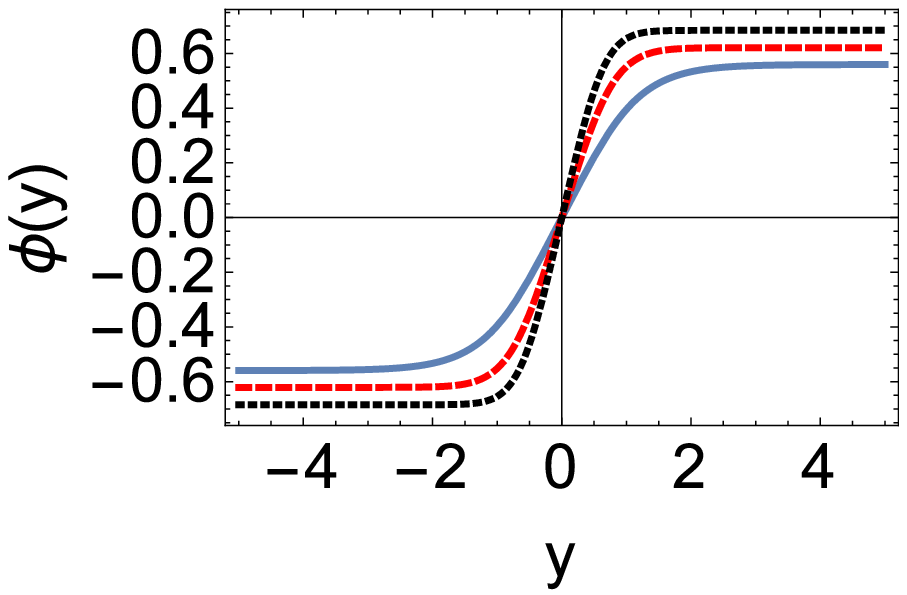}}
\subfigure[$V(y)$]{\includegraphics[width=4cm]{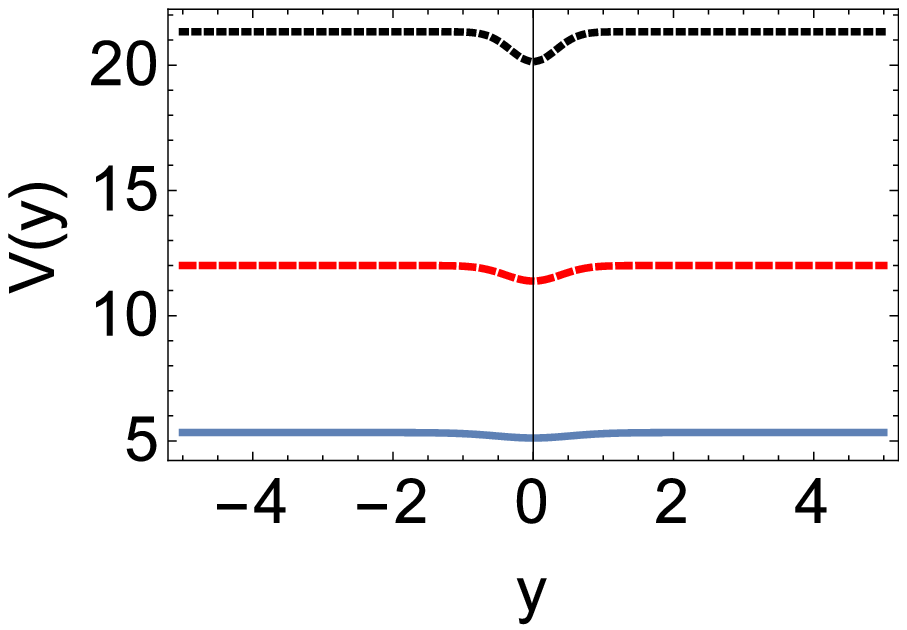}}
\subfigure[$\rho(y)$]{\includegraphics[width=4cm]{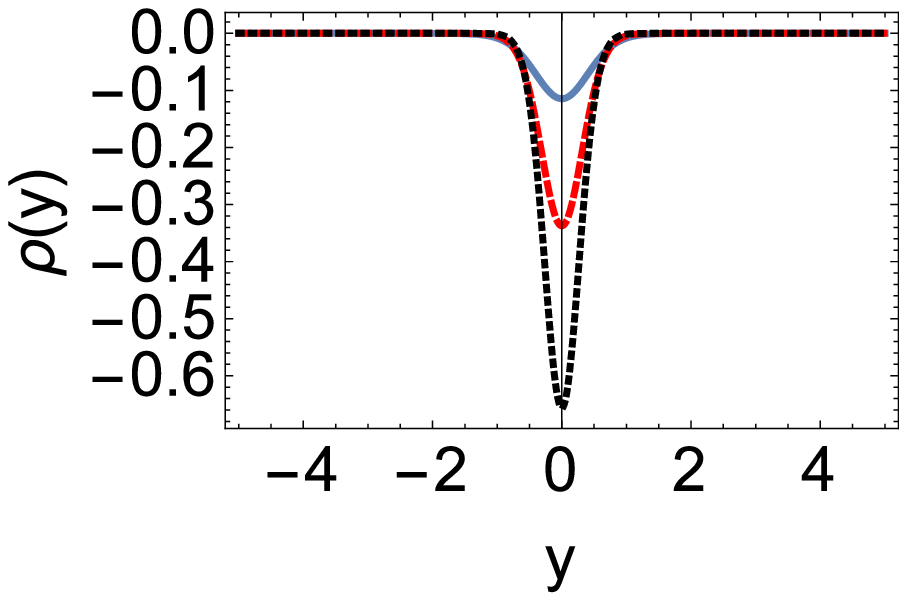}}
\subfigure[$R(y)$]{\includegraphics[width=4cm]{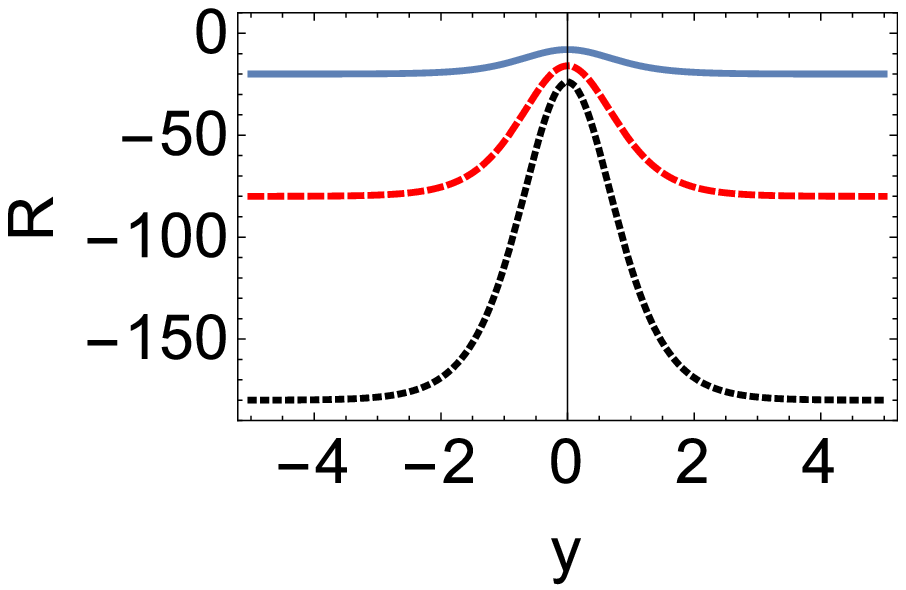}}
}
\caption{The shapes of the thick brane solution. (a) The warp factor $a(y)$. (b) The scalar field $\phi(y)$. (c) The scalar potential $V(y)$. (d) The energy density $\rho(y)$. (e) The curvature scalar $R(y)$. The parameters are set to $\kappa=1$, $k=1$, $\lambda=1$, along with $n=5/2$ (blue solid line), $n=9/2$ (red dashed line), and $n=13/2$ (black dotted line).
}\label{brhoR}
\end{figure}

Next, we discuss the behavior of the scalar potential. By using Eq.~\eqref{seom1}, we have the following relations
\begin{eqnarray}
V_{\phi}&=&\frac{4a'\phi'}{a}+\phi'',\label{Vp1}\\
V_{\phi\phi}&=&\frac{4aa'\phi''+4aa''\phi'^{2}-4a'^{2}\phi'^{2}}{a^{2}\phi'}+\frac{\phi'''}{\phi'},\label{Vp2}
\end{eqnarray}
where $V_{\phi\phi}$ represents the second derivative of the scalar potential with respect to the scalar field. It is obvious that the concrete forms of $V_{\phi}$ and $V_{\phi\phi}$ with respect to $y$ can be obtained by using Eqs.~\eqref{a1} and~\eqref{phi1}. Since the scalar field $\phi$ is an odd function and monotonously increases with the extra dimension $y$, one can conclude that when $\phi=0$, $V_{\phi}|_{\phi=0}=0$ and $V_{\phi\phi}|_{\phi=0}=k^2 \left(3\times2^{3/4} k K^{3/4} (2 n-1) (2 n+3)^{3/4}+2 \kappa  (3-2 n) n^{1/4}\right)\\/(12 n^{1/4} \kappa)$. Hence, with a proper value of $k$, we can have $V_{\phi\phi}|_{\phi=0}>0$, which denotes a minimum of the scalar potential at $\phi=0$. When $y$ approaches to infinity, or equivalently $\phi=\pm v_{0}$, one can also find that $V_{\phi}|_{\phi=\pm v_{0}}=0$. It is hard to decide whether $\phi=\pm v_{0}$ are the extreme points because of the cutoff on the scalar field. Nevertheless, as shown in Fig.~\ref{brhoR}, by comparing the values of the scalar potential $V(\phi=0)$ and $V(\phi=\pm v_{0})$, we find that there exists a vacuum at the origin of the extra dimension corresponding to the lowest energy state of the scalar field.

To check whether the background solution corresponds to a thick braneworld model, we should discuss the energy density of the brane. As we know, in most braneworld models~\cite{Zhou1,Zhong1}, the vacuum of the scalar field is usually located at the boundary of the extra dimension, and the energy density of the brane is contributed by the scalar field only. So, one should subtract the contribution from the cosmological constant, which may exist in the scalar potential and may lead to a nonzero vacuum energy density, to fix the energy density to be zero at the boundary. However, this is usually not the case in modified gravities. Because once we try to modify the Einstein-Hilbert action, we are actually including some unknown effects on the geometry. Hence, it is acceptable to change the scalar potential while keeping some of the others unchanged~\cite{Gu1}. In this paper, the vacuum is no longer located at the boundary but at the origin of the extra dimension. Therefore, instead of subtracting the nonzero vacuum energy density, we find that it is important to drop a constant term, $-\frac{\lambda}{b\kappa^{2}}$, in the energy density to fix it to be zero far from the brane. Further, for a static observer with four-velocity $\omega^{M}$, the brane's energy density is defined as $\rho=T_{MN}\omega^{M}\omega^{N}+\frac{\lambda}{b\kappa^{2}}$ and has the following explicit expression
\begin{eqnarray}
\rho(y)&=&-T_{~0}^{0}+\frac{\lambda}{b\kappa^{2}} \nonumber \\
       &=& -\frac{   (2 n-1)^{5 n+\frac{3}{2}} (2 n+3)^{\frac{3}{2}-n} k^2}
                 {3\times 4^{2 n+1} n^{4 n+1} \kappa ^2}
            \text{sech}^{2 n+1}(k y).\nonumber\\
            &~&
\end{eqnarray}
Since, as we have mentioned above, the parameter $n$ should be larger than $1/2$, the energy density does not diverge. It is obvious that, the behavior of the energy density denotes the localization of the thick brane near the origin of the extra dimension with the brane's thickness determined by $k$. Moreover, the brane can hide its thickness from low-energy tests on the brane by enlarging $k$. Note that, though we only consider some explicit cases of the background spacetime in this section and Sec.~\ref{41}, one could check that some fundamental properties of the brane is unchanged in the case of $n>1/2$.

After the analysis of the localization of the thick brane, we should further investigate the effects of the thick brane on the geometry of five-dimensional spacetime.         Using the above thick brane solution and the definition of the physical curvature scalar, $R\equiv g^{MN}R_{MN}(g)$, one obtains
\begin{equation}
R=-\frac{1}{4} k^2 (2 n-1) \left[(10 n-13) \tanh ^2(k y)+8\right].
\end{equation}
It is obvious that, when $n>1/2$, the physical curvature scalar of the bulk spacetime approaches to a negative constant, i.e., $2k^2 (1-2n)$, far from the brane, which implies an asymptotically anti$-$de Sitter (AdS) spacetime. The extremum of the physical curvature scalar implies that, being consistent with the configuration of the thick brane, the matter is mainly distributed on the brane. The shapes of the thick brane solution, energy density, and physical curvature scalar are shown in Fig.~\ref{brhoR}.

\section{Metric perturbations}\label{4}

\subsection{Localization of the massless graviton}\label{41}

In this subsection, we investigate the metric perturbations and the localization of the massless graviton. With great interest in the tensor perturbation, we use the transverse-traceless (TT) gauge to decouple the vector perturbations and the scalar perturbations from the tensor perturbation in spacetime metric, as done in Ref.~\cite{KeYang}. Thus, the perturbed metrics should be
\begin{eqnarray}
ds^{2}&=&g_{MN}dx^{M}dx^{N}=(\bar{g}_{MN}+\bigtriangleup g_{MN})dx^{M}dx^{N}\nonumber\\
&=&a^{2}(y)(\eta_{\mu\nu}+h_{\mu\nu})dx^{\mu}dx^{\nu}+dy^{2},\\
d\tilde{s}^{2}&=&q_{MN}dx^{M}dx^{N}=(\bar{q}_{MN}+\bigtriangleup q_{MN})dx^{M}dx^{N}\nonumber\\
&=&u(y)(\eta_{\mu\nu}+\gamma_{\mu\nu})dx^{\mu}dx^{\nu}+2\gamma_{\mu5}dx^{\mu}dy\nonumber\\
&~&+v(y)(1+\gamma_{55})dy^{2},
\end{eqnarray}
with
\begin{equation}
\bar{g}_{MN}=\left(
\begin{array}{cc}
a^{2}\eta_{\mu\nu}&0\\
0&1\\
\end{array}
\right),~~
\triangle g_{MN}=\left(
\begin{array}{cc}
a^{2}h_{\mu\nu}&0\\
0&0\\
\end{array}
\right),
\end{equation}
and
\begin{equation}
\bar{q}_{MN}=\left(
\begin{array}{cc}
u\eta_{\mu\nu}&0\\
0&v\\
\end{array}
\right),~~
\triangle q_{MN}=\left(
\begin{array}{cc}
u\gamma_{\mu\nu}&\gamma_{\mu5}\\
\gamma_{\mu5}&v\gamma_{55}\\
\end{array}
\right),
\end{equation}
where the notations $\gamma$ and $h$, respectively, represent the perturbations of the auxiliary and spacetime metrics, $\bar{g}_{MN}$ and $\bar{q}_{MN}$ are the background metrics. From now on, we will mainly focus on the TT part of $h$, which satisfies the TT gauge condition
\begin{equation}\label{TTh}
\partial_{\rho}(h^{\text{TT}})_{~\mu}^{\rho}=0=(h^{\text{TT}})_{~\rho}^{\rho}\equiv h^{\text{TT}}.
\end{equation}
To obtain the inverse of the metrics, one should use the relations $g_{MN}g^{MK}=\delta_{~N}^{K}$ and $q_{MN}q^{MK}=\delta_{~N}^{K}$. Moreover, aiming to obtain the linear perturbation equations, we drop the high-order perturbation terms in fields for convenience. Finally, one has
\begin{eqnarray}
\triangle^{(1)}g^{PK}\equiv\delta g^{PK} &=& -\bar{g}^{NP}\bar{g}^{MK}\triangle g_{MN}, \\
\triangle^{(1)}q^{PK}\equiv\delta q^{PK} &=& -\bar{q}^{NP}\bar{q}^{MK}\triangle q_{MN}.
\end{eqnarray}
Note that, here and after, we use the notation $\delta$ together with a tensor to denote its first-order perturbation. The inverse of metrics can be expressed as
\begin{eqnarray}
g^{MN}&=&\left(
\begin{array}{cc}
a^{-2}[\eta^{\mu\nu}-(h^{\text{TT}})^{\mu\nu}]&0\\
0&1\\
\end{array}
\right),\\
q^{MN}&=&\left(
\begin{array}{cc}
u^{-1}(\eta^{\mu\nu}-\gamma^{\mu\nu})&-u^{-1}v^{-1}\gamma_{~5}^{\mu}\\
-u^{-1}v^{-1}\gamma_{~5}^{\mu}&v^{-1}(1-\gamma_{55})\\
\end{array}
\right).
\end{eqnarray}
For the sake of simplicity, we give the expressions of the perturbed auxiliary tensor $\Omega_{~M}^{N}$ and its determinant as follows
\begin{eqnarray}
\Omega_{~M}^{N}&\approx&\bar{\Omega}_{~M}^{N}+\delta\Omega_{~M}^{N},\\
\big|\hat{\Omega}\big|&\approx&\big|\bar{\hat{\Omega}}\big|+\delta\big|\hat\Omega\big|,
\end{eqnarray}
where $\delta\Omega_{~M}^{N}$ and $\delta\big|\hat\Omega\big|$ can be obtained from the definition of $\Omega_{~M}^{N}$. Furthermore, we should use the equations of motion~\eqref{am4} and~\eqref{Elf2} to obtain the linear perturbation equations
\begin{eqnarray}
\delta g_{MN}+b\,\delta R_{MN}&=&2^{-\frac{2}{3}}|\bar{\hat{\Omega}}|^{-\frac{1}{3}}f_{|\bar{\hat{\Omega}}|}^{-\frac{2}{3}}\delta q_{MN}\nonumber\\
&~&-\frac{1}{3}2^{\frac{1}{3}}|\bar{\hat{\Omega}}|^{-\frac{1}{3}}f_{|\bar{\hat{\Omega}}|}^{-\frac{5}{3}}f_{|\bar{\hat{\Omega}}|,|\bar{\hat{\Omega}}|}\bar{q}_{MN}\delta|\hat{\Omega}|\nonumber\\
&~&-\frac{1}{3}2^{-\frac{2}{3}}|\bar{\hat{\Omega}}|^{-\frac{4}{3}}f_{|\bar{\hat{\Omega}}|}^{-\frac{2}{3}}\bar{q}_{MN}\delta|\hat{\Omega}|,\label{pam1}\\
-b\kappa^{2}\delta T^{MN}&=&f_{|\bar{\hat{\Omega}}|}\bar{g}^{MN}\delta\big|\hat{\Omega}\big|+\big(f-\lambda\big)\delta g^{MN}\nonumber\\
&~&+\frac{1}{3}2^{\frac{11}{3}}\big|\bar{\hat{\Omega}}\big|^{\frac{1}{3}}f_{|\bar{\hat{\Omega}}|}^{\frac{5}{3}}\bar{q}^{MN}\delta\big|\hat{\Omega}\big|\nonumber\\
&~&+\frac{5}{3}2^{\frac{5}{3}}|\bar{\hat{\Omega}}|^{\frac{4}{3}}f_{|\bar{\hat{\Omega}}|}^{\frac{2}{3}}f_{|\bar{\hat{\Omega}}|,|\bar{\hat{\Omega}}|}\bar{q}^{MN}\delta|\hat{\Omega}|\nonumber\\
&~&+2^{\frac{5}{3}}|\bar{\hat{\Omega}}|^{\frac{4}{3}}f_{|\bar{\hat{\Omega}}|}^{\frac{5}{3}}\delta q^{MN}-2f_{|\bar{\hat{\Omega}}|}\bar{g}^{MN}\delta|\hat{\Omega}|\nonumber\\
&~&-2|\bar{\hat{\Omega}}|f_{|\bar{\hat{\Omega}}|,|\bar{\hat{\Omega}}|}\bar{g}^{MN}\delta|\hat{\Omega}|\nonumber\\
&~&-2|\bar{\hat{\Omega}}|f_{|\bar{\hat{\Omega}}|}\delta g^{MN}.\label{pElf1}
\end{eqnarray}
It is important to note that, as discussed in Ref.~\cite{KeYang}, the TT components of the perturbed metrics are decoupled with the scalar field perturbation. So, in this paper, we only consider $h$ in the perturbed energy-momentum tensor:
\begin{eqnarray}\label{dT}
\delta T^{MN}&=&-b\kappa^{2}\Big(\delta g^{MP}\bar{g}^{NK}\partial_{P}\phi\partial_{K}\phi\nonumber\\
&~&+\bar{g}^{MP}\delta g^{NK}\partial_{P}\phi\partial_{K}\phi-\frac{1}{2}\delta g^{MN}\bar{g}^{PK}\partial_{P}\phi\partial_{K}\phi\nonumber\\
&~&-\frac{1}{2}\bar{g}^{MN}\delta g^{PK}\partial_{P}\phi\partial_{K}\phi-\delta g^{MN}V\Big).
\end{eqnarray}
By calculating Eqs.~\eqref{pElf1} and~\eqref{dT}, one can obtain the following relations:
\begin{equation}\label{TTampm}
\gamma_{\mu\nu}=h_{\mu\nu}^{\text{TT}},~~
\gamma_{\mu5}=0,~~
\gamma_{55}=0.
\end{equation}
It is interesting that the above relations~\eqref{TTampm} are the same as the ones obtained in EIBI theory~\cite{KeYang}. Actually, before the calculation of the above relations, as shown in Eq.~\eqref{pam1}, we expected that the perturbations on the metrics would be related by the function $f$. But it seems that, once we use the TT gauge on the spacetime metric perturbations, no matter what kinds of $f$ we employ, the auxiliary metric perturbations are always decoupled from $f$ and equal to the TT components of the spacetime metric perturbations. It means that, once we use the TT gauge condition on the spacetime metric perturbations, we are actually using the same gauge condition on the auxiliary metric perturbations. Note that the above relations will lead to $\delta|\hat{\Omega}|=0$, and one can further simplify Eq.~\eqref{pam1} as
\begin{equation}\label{pam3}
\square^{(4)}h_{\mu\nu}^{\text{TT}}+\frac{u}{v}(h_{\mu\nu}^{\text{TT}})''+\Big(\frac{2u'}{v}-\frac{uv'}{2v^{2}}\Big)(h_{\mu\nu}^{\text{TT}})'=0,
\end{equation}
which can be reduced to the result given in EIBI theory~\cite{QiMing} if one sets $f(|\hat{\Omega}|)=|\hat{\Omega}|^{1/2}$. Further, one can use the coordinate transformation $dy=\sqrt{u(y)/v(y)}dz$ and the decomposition of the $n'$th mode of the tensor perturbation
\begin{equation}
h_{\mu\nu}^{\text{TT}(n')}(x^{\rho},z)=\epsilon_{\mu\nu}^{(n')}(x^{\rho})u(z)^{-3/4}\psi_{n'}(z)
\end{equation}
to obtain the Schr\"{o}dinger-like equation of the $n'$th graviton KK mode. The fifth component of Eq.~\eqref{pam3} is
\begin{equation}\label{Sl1}
-\partial_{z}^2\psi_{n'}(z)+U(z)\psi_{n'}(z)=m_{n}^{2}\psi_{n'}(z)
\end{equation}
with the effective potential $U(z)$ given by
\begin{equation}
U(z)=-\frac{3}{16}\frac{\big(\partial_{z}u\big)^{2}-4u\partial_{z}^2 u}{u^{2}}.
\end{equation}
The four-dimensional mass of the $n'$th graviton KK mode, $m_{n'}$, is defined by
\begin{equation}
\square^{(4)}\epsilon_{\mu\nu}^{(n')}(x^{\mu})=m_{n'}^{2}\epsilon_{\mu\nu}^{(n')}(x^{\mu}).
\end{equation}
The Schr\"{o}dinger-like equation~\eqref{Sl1} implies that, for the case of an infinite extra dimension, the unbound graviton KK modes can propagate in the extra dimension while the bound ones are trapped in the effective potential. Moreover, the effective potential, which results from the warped extra dimension and usually acts like a potential well, affects the behavior of the KK gravitons. In general, the graviton zero mode  is usually bounded in the potential well while the massive graviton is unbounded. Note that the factorization of Eq.~\eqref{Sl1},
\begin{equation}
\bigg(\frac{d}{dz}+\frac{3}{4}\frac{\partial_{z}u}{u}\bigg)\bigg(-\frac{d}{dz}+\frac{3}{4}\frac{\partial_{z}u}{u}\bigg)\psi_{n'}=m_{n'}^{2}\psi_{n'},
\end{equation}
ensures that the mass square of every graviton KK mode cannot be negative, which highlights the absence of tachyon states and the stability of the tensor perturbations. Here, we are interested in the behavior of the massless graviton whose localization on the thick brane is an important indicator of reconstruction of the four-dimensional Newtonian potential. The corresponding equation of the massless graviton is given as follows:
\begin{equation}\label{Sl2}
\bigg(\frac{d}{dz}+\frac{3}{4}\frac{\partial_{z}u}{u}\bigg)\bigg(\frac{d}{dz}-\frac{3}{4}\frac{\partial_{z}u}{u}\bigg)\psi_{0}=0.
\end{equation}
Then, one can easily obtain the function of the zero mode KK graviton:
\begin{equation}\label{zg}
\psi_{0}=Au^{3/4}.
\end{equation}
Using the normalization condition $\int_{z_{\text{min}}}^{z_{\text{max}}}\psi_{0}^{2}dz=1$, one can further determine the normalization coefficient as $A^{-2}=\int_{-\infty}^{+\infty}uv^{1/2}dy$.

\begin{figure}[!htb]
\center{
\subfigure[$U(y)$]{\includegraphics[width=4cm]{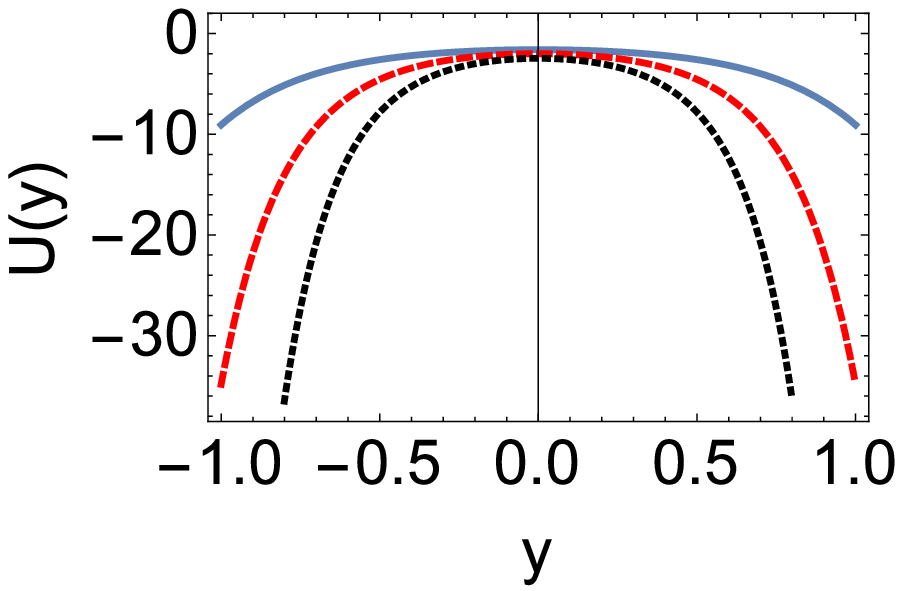}}
\subfigure[$\psi_{0}(y)$]{\includegraphics[width=4cm]{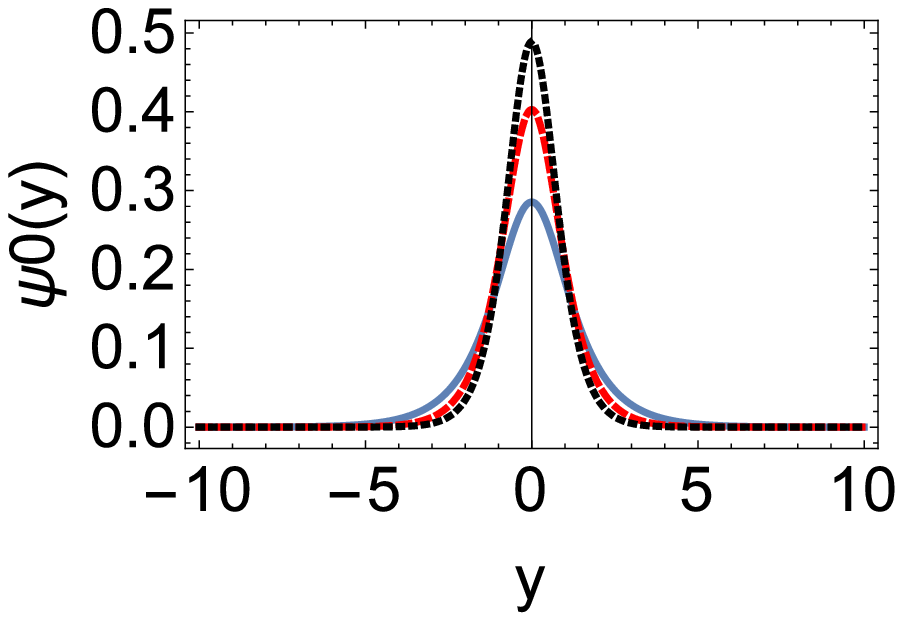}}
\subfigure[$z(y)$]{\includegraphics[width=4cm]{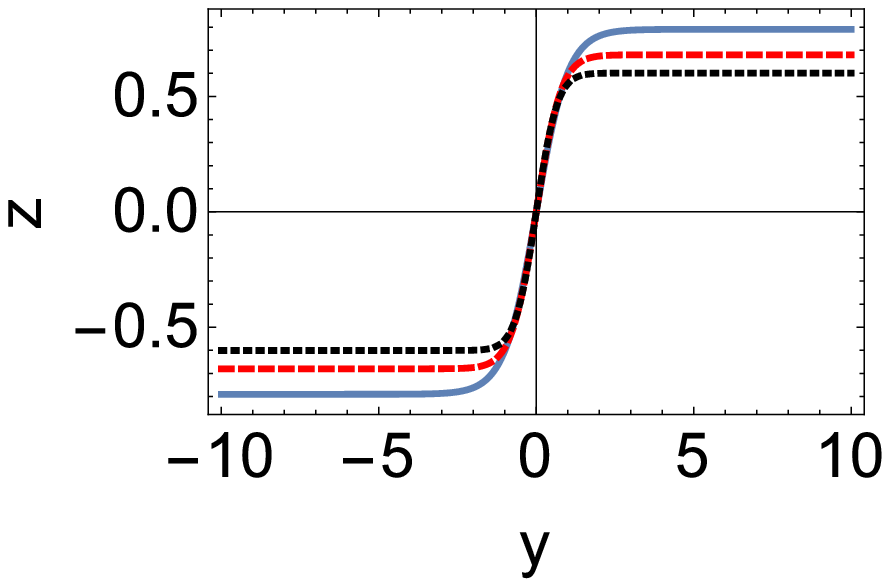}}
\subfigure[$\psi_{0}(z)$]{\includegraphics[width=4cm]{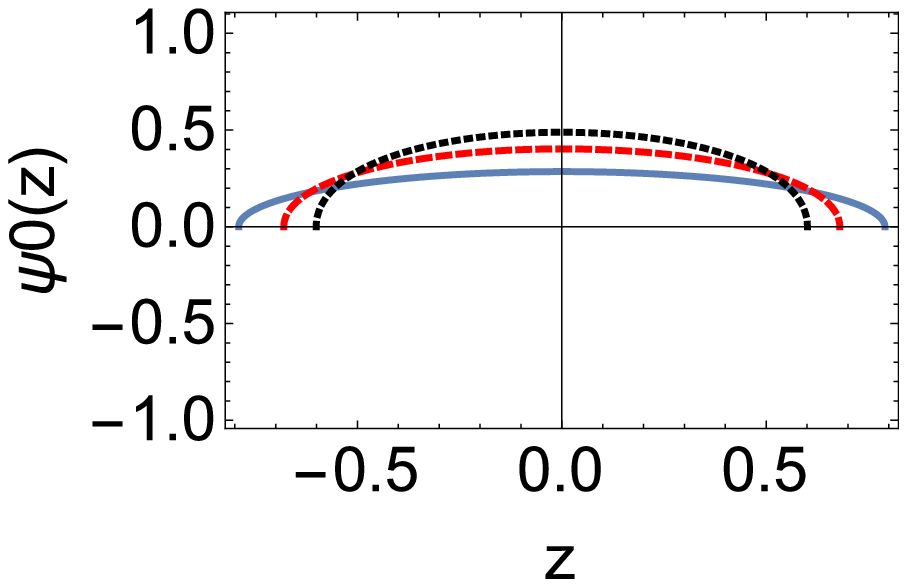}}
}
\caption{The shapes of the effective potential, the massless graviton, and the function $z(y)$. (a) The effective potential $U(y)$. (b) The massless graviton $\psi_{0}(y)$ in the $y$ coordinate. (c) The function $z(y)$. (d) The massless graviton $\psi_{0}(z)$ in the $z$ coordinate. The parameter $n$ is set to $n=5/2$ (blue solid line), $9/2$ (red dashed line), and $13/2$ (black dotted line), and the parameters $\kappa$, $k$, and $\lambda$ are set to unity for convenience.}\label{zpsiU}
\end{figure}

In order to recover the four-dimensional Newtonian potential, we investigate the localization of the massless graviton. The corresponding analytical function of the massless graviton is given as
\begin{eqnarray}
\psi_{0}(y)=AK^{\frac{3}{4}}\text{sech}^{\frac{3}{8}+\frac{n}{4}}(ky),
\end{eqnarray}
where the parameter $K$ is given in Eq.~\eqref{K1} and the normalization coefficient reads $A^{-2}=\frac{K^{3/2}\sqrt{2n\pi}}{k\sqrt{3+2n}}\Gamma(\frac{3}{4}+\frac{n}{2})/\Gamma(\frac{5}{4}+\frac{n}{2})$. The shapes of the effective potential, the massless graviton, and the function $z(y)$ are shown in Fig.~\ref{zpsiU}. We find that the effective potential is infinite at the boundary of the extra dimension, and hence, the ground state of gravitons should vanish at the boundary. Therefore, the massless graviton is localized near the brane. On the other hand, we find that, with the solution~\eqref{a1}-\eqref{v1}, the coordinate transformation leads to a finite extra-dimensional coordinate $z$. Unlike the infinite one obtained in Refs.~\cite{KeYang,QiMing}, this finite extra-dimensional coordinate makes all the gravitons bounded in the extra dimension, and the shapes of gravitons are determined by the effective potential.

\subsection{Four-dimensional low-energy effective theory}

In this subsection, we investigate the low-energy effective theory of a specific family of theories, i.e., $f(|\hat{\Omega}|)=|\hat{\Omega}|^{\frac{1}{2}+n}$, on the brane. With a small value of $bR_{MN}$, by expanding the action~\eqref{ea} to second order in $b$, the action reads
\begin{eqnarray}\label{ap}
S&\approx&\frac{1}{2\tilde{\kappa}^{2}}\int d^{D}x\sqrt{-g}\Big[\tilde{R}-2\Lambda_{\text{eff}}+\frac{(2n+1)}{4}b\tilde{R}^{2}\nonumber\\
&~&-\frac{b}{2}R_{MN}R^{MN}+\mathcal{O}(b^{2})\Big]+S_{m}(g_{MN},\phi),
\end{eqnarray}
where $\Lambda_{\text{eff}}=\frac{\lambda-1}{b(2n+1)}$ is the effective cosmological constant, $\tilde{\kappa}^{2}=\frac{(2n+1)\kappa^{2}}{2}$, and $\tilde{R}=g^{MN}R_{MN}(\Gamma)$. Recalling Eq.~\eqref{am2} and expanding the determinant of the auxiliary tensor $\Omega_{~N}^{M}$, the relation between the auxiliary metric and the spacetime metric is given as follows
\begin{eqnarray}
q_{MN}&\approx&(1+2n)^{\frac{2}{D-2}}g_{MN}\nonumber\\
&~&+b(1+2n)^{\frac{2}{D-2}}\Big(R_{MN}+\frac{2n}{D-2}g_{MN}\tilde{R}\Big)\nonumber\\
&~&+\mathcal{O}(b^{2}).
\end{eqnarray}
It means that, with a lowest-order approximation, the independent connection is also compatible with the spacetime metric. In other words, the action~\eqref{ap} will reproduce the Einstein-Hilbert action together with the effective cosmological constant in the low-energy approximation.
%What's more, with a tiny parameter $n$, the auxiliary metric approximates to the spacetime metric in low-energy regime.
After considering the TT gauge condition on the metric perturbations and using thick brane solutions, one can obtain a four-dimensional effective theory in {\color{blue} low-energy regime}. Also, neglecting the contributions of the effective cosmological constant and matter for convenience, the four-dimensional effective action in the low-energy approximation is finally given as follows
\begin{equation}\label{ap2}
S\supset\frac{1}{2 \kappa_4^{2}}\int{d^{4}x\sqrt{-|g^{(4)}_{\mu\nu}|}R^{(4)}},
\end{equation}
where $g^{(4)}_{\mu\nu}(x^{\rho})$ is the four-dimensional metric on the brane. The four-dimensional curvature scalar is now defined as $R^{(4)}(x^{\rho})=g^{(4)\mu\nu}(x^{\rho})R^{(4)}_{\mu\nu}(g^{(4)}_{\lambda\rho})$. Moreover, the integral of the action over the extra dimension is absorbed in the constant $\kappa_4$ with
\begin{eqnarray}
%\kappa_4^{2}&=&\frac{2^{4n}n^{4n+2}(2n+3)^{n+1/2}}{(2n-1)^{5(n+1/2)}}bk\kappa^{2}.
\kappa^{2}&=&\kappa_{4}^{2}\,\frac{2n+1}{k}\,D_{1}(y)|^{+\infty}_{-\infty}\nonumber\\
&=&\kappa_{4}^{2}\sqrt{\pi }\,\frac{2n+1}{k}\frac{\Gamma \left(n-\frac{1}{2}\right)}{\Gamma (n)},
\end{eqnarray}
where
\begin{eqnarray}
D_{1}(y)&=&\sinh (k y)\,_2F_1\left(\frac{1}{2},n;\frac{3}{2};-\sinh ^2(k y)\right).
\end{eqnarray}
Further, the ratio between the $D$-dimensional and the four-dimensional Newtonian gravitational constants is
\begin{eqnarray}
\frac{G}{G_{4}}&=&\frac{\lambda_{1}(n)}{k},\label{G1}\\
\lambda_{1}(n)&=&\sqrt{\pi }\,(2n+1)\frac{\Gamma \left(n-\frac{1}{2}\right)}{\Gamma (n)}.\label{lambda1}
\end{eqnarray}
Hence, the effective four-dimensional (reduced) Planck scale of this low-energy effective theory should be $M_{Pl}=\kappa_4^{-1}\sim10^{19}\,\text{GeV}$, while the fundamental five-dimensional mass scale is $M_{*}=\kappa^{-2/3}$.

\subsection{Correction to Newtonian gravitational potential}

To give a prediction on the correction to the effective Newtonian gravitational potential in this $f(|\hat{\Omega}|)$ theory, we should investigate the contribution from the massive gravitons. As we have mentioned above, a finite extra-dimensional coordinate $z$ is finally obtained. Thus, in a qualitative analysis, the KK spectrum of gravitons is discrete and the function of the $n'$th graviton KK mode should be given as
\begin{eqnarray}
\psi_{n'}(z)\sim A_{n'}\text{sin}\frac{\pi n'}{L}z+B_{n'}\text{cos}\frac{\pi n'}{L}z,
\end{eqnarray}
where $A_{n'}=\sqrt{2/L}$ and $B_{n'}=\sqrt{2/L}$ are the normalization coefficients for the odd and even modes, respectively, and the size $L$ of the extra dimension in the conformal coordinate $z$ can be calculated from the coordinate transformation $dz =\sqrt{v(y)/u(y)}~ dy$ as
\begin{eqnarray}
L=\frac{\lambda_{2}(n)}{k},\label{L1}
\end{eqnarray}
where
\begin{eqnarray}
\lambda_{2}(n)&=&\sqrt{\frac{2n}{2 n+3}}D_{2}(y)|_{-\infty}^{+\infty}\nonumber\\
 &=&\sqrt{\frac{2n\pi}{2 n+3}}\,\frac{\Gamma \left(\frac{n}{4}+\frac{3}{8}\right)}{\Gamma \left(\frac{n}{4}+\frac{7}{8}\right)},\label{lambda2}\\
D_{2}(y)&=&\sinh (k y) \, _2F_1\left(\frac{1}{2},\frac{2 n+7}{8} ;\frac{3}{2};-\sinh ^2(k y)\right).\nonumber\\
&~&
\end{eqnarray}
Also, the contribution from the massive gravitons to the four-dimensional effective Newtonian potential on our brane at $z=0$ will only come from the even modes. Then, we conclude that the gravitational potential between two masses on the brane is~\cite{Bazeia2,RS,KeYang3}
\begin{eqnarray}
V_{\text{eff}}(r)&\approx & G_4\frac{m'_{1}m'_{2}}{r}+\sum_{n'=1}^{\infty}G\frac{m'_{1}m'_{2}e^{-m_{n'}r}}{r}|\psi_{n'}(0)|^{2} \nonumber \\
&=& G_4\frac{m'_{1}m'_{2}}{r}\bigg[1+\frac{G}{G_{4}} \frac{2}{L} \frac{1}{e^{\pi r/L}-1}\bigg],
\end{eqnarray}
where $m'_{1}$ and $m'_{2}$ are the mass of the particles, $G_{4}$ is the four-dimensional Newtonian gravitational constant, and $m_{n'}=n'\pi/L$ is the mass of the $n'$th graviton KK mode. It is now clear that the first term above is contributed from the massless graviton and that massive gravitons will give some corrections to the effective Newtonian potential. Moreover, the correction term expressed by the exponential function will vanish in a large scale, and then recovers the usual Newtonian potential between the particles. On the other hand, if the distance between particles is short enough ($r \ll L$), the total correction contributed from the massive gravitons cannot be neglected. For this case, the effective Newtonian potential will be
\begin{eqnarray}
V_{\text{eff}}(r)\approx G_4\frac{m'_{1}m'_{2}}{r}\bigg[1+\frac{2\lambda_{1}}{\pi k r} \bigg],\label{NVeff}
\end{eqnarray}
where we have used Eq.~\eqref{G1}. The recent test of the gravitational inverse-square law showed that, at a $95\%$ confidence level, the usual Newtonian potential will hold down to a length scale at $59\,\mu\text{m}$~\cite{Jun1}. It means that the contribution from the second term of the effective Newtonian potential~\eqref{NVeff} should be negligible at $r\gtrsim 59\,\mu\text{m}$.
Based on this fact, we give a strong constraint on the parameter $k$ in $f(|\hat{\Omega}|)$ theory:
\begin{eqnarray}
k&>&\frac{2\lambda_{1}}{\pi r_{c}},\label{k1}
\end{eqnarray}
where we assume $r_{c}=59\,\mu\text{m}$ being the critical distance of breaking the gravitational inverse-square law between the masses.
\begin{figure}[!htb]
\center{
\subfigure[$\lambda_{1}(n)$]{\includegraphics[width=4cm]{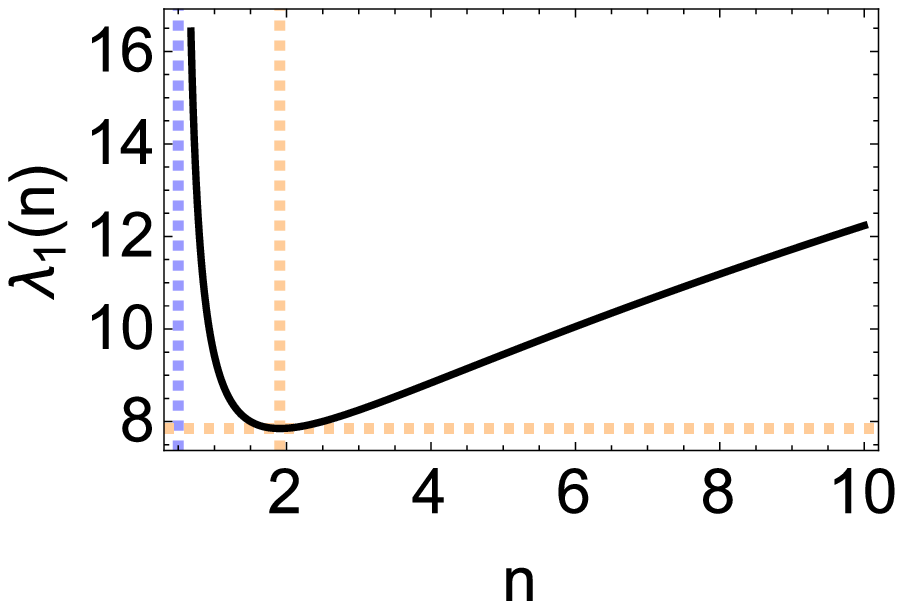}}
\subfigure[$\lambda_{2}(n)$]{\includegraphics[width=4cm]{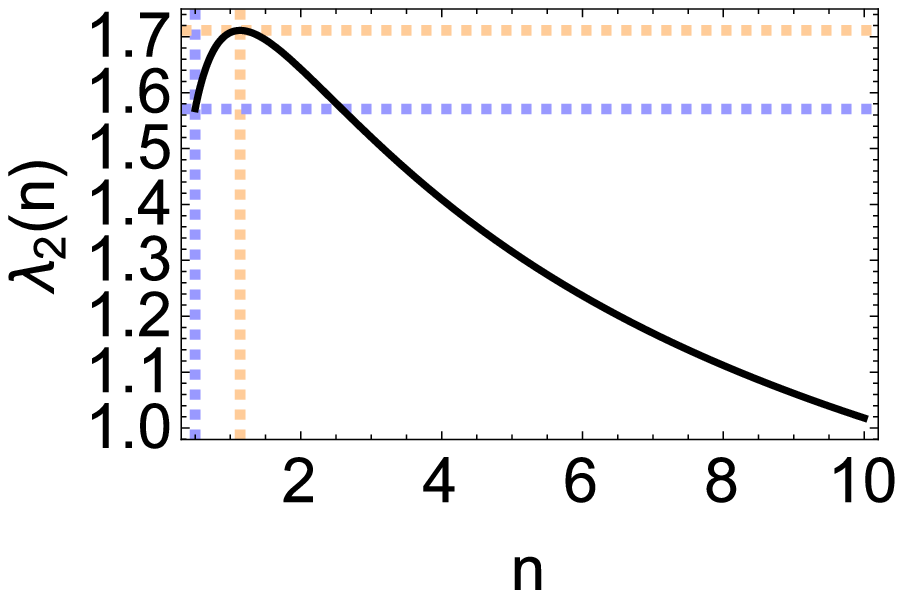}}
\subfigure[$\lambda_{3}(n)$]{\includegraphics[width=4cm]{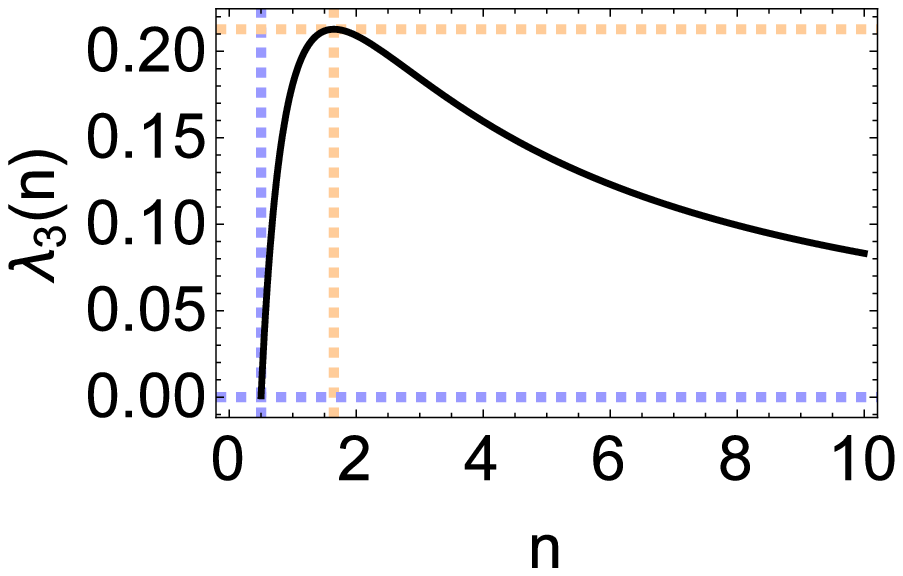}}
\subfigure[$\lambda_{4}(n)$]{\includegraphics[width=4cm]{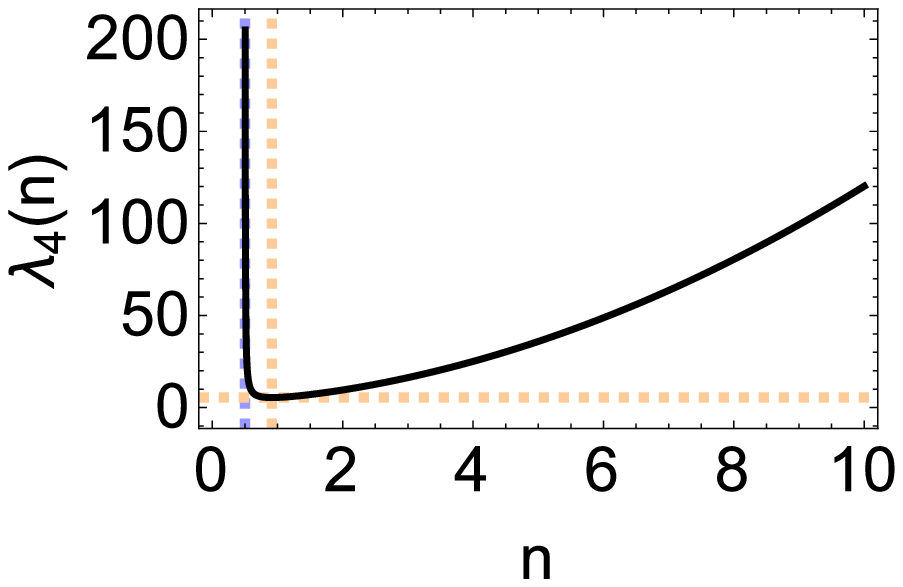}}
}
\caption{The relation between the parameters $\lambda_{i}$ and $n$. (a) $\lambda_{1}(n)$ defined in Eq.~\eqref{lambda1}. (b) $\lambda_{2}(n)$ defined in Eq.~\eqref{lambda2}. (c) $\lambda_{3}(n)$ defined in Eq.~\eqref{lambda3}. (d) $\lambda_{4}(n)$ defined in Eq.~\eqref{lambda4}. The blue dashed lines denote the value of the parameters with $n$ approaching to $1/2$, while the orange dashed lines localize the extreme points of the parameters.}\label{lambda1234}
\end{figure}
As is shown in Fig.~\ref{lambda1234}, the parameter $\lambda_{1}$ has a minimum value $\lambda_{1\,\text{min}}=7.84871$ at the point $n=1.90554$. Therefore, $k$ should have a mass scale of at least $10^{-2}\,$eV and will increase with the deviation of $n$ from that point. Then, the constraint on the fundamental mass scale can be obtained by using Eq.~\eqref{G1}:
\begin{eqnarray}
M^{3}_{*}&>&\frac{2}{\pi r_{c}}M^{2}_{Pl}.
\end{eqnarray}
It means that the fundamental mass scale should be at least $10^{5}\,\text{TeV}$. Recalling Eqs.~\eqref{L1} and~\eqref{k1}, one further gives the constraint on the size of the extra dimension, $L$, in the conformal coordinate $z$ as
\begin{eqnarray}
L&<&\frac{\pi r_{c}}{2}\lambda_{3}(n),\label{L2}
\end{eqnarray}
where
\begin{eqnarray}
\lambda_{3}(n)=\frac{\lambda_{2}}{\lambda_{1}}.\label{lambda3}
\end{eqnarray}
From Fig.~\ref{lambda1234}, one finds that when $n$ approaches to $1/2$, the parameter $\lambda_{3}$ approaches to zero. Also, $\lambda_{3}$ gets its maximal value, $\lambda_{3\,\text{max}}=0.212763$, at the point $n=1.64715$, which implies that $L$ has a length scale of at most $0.1\,\text{mm}$. The mass of the $n'$th graviton KK mode and Eq.~\eqref{L2} give another constraint:
\begin{eqnarray}
m_{n'}>\frac{2}{r_{c}}\lambda_{4}(n),
\end{eqnarray}
with
\begin{eqnarray}
\lambda_{4}(n)=\frac{n}{\lambda_{3}}.\label{lambda4}
\end{eqnarray}
It is obvious that the parameter $\lambda_{4}$ has a minimum value, $\lambda_{4\,\text{min}}=5.45515$, when $n=0.912792$, and one concludes that the mass spectrum of the massive gravitons is at least $10^{-2}\,\text{eV}$. Note that the value of the mass gap is dependent on the value of $\lambda_{4}$, and, as shown in Fig.~\ref{lambda1234}, it will increase to infinity when $n$ approaches to infinity and $1/2$. In addition, we only give the lower limit of the fundamental mass scale, which means that the coupling strength of each massive graviton to matter is about $1/M_{*}\sim10^{-5}\,\text{TeV}^{-1}$ at most. However, one could always assume that the fundamental mass scale is much larger than that limit, for example, $M_{*}\sim10^{16}\,\text{TeV}$. Therefore, the effect of each massive graviton on particle physics can be negligible, and then one should count the combined effect of all the massive gravitons to give an observable correction to both the usual Newtonian potential and particle physics.

\begin{figure}[!htb]
\center{
\subfigure[$k$]{\includegraphics[width=4cm]{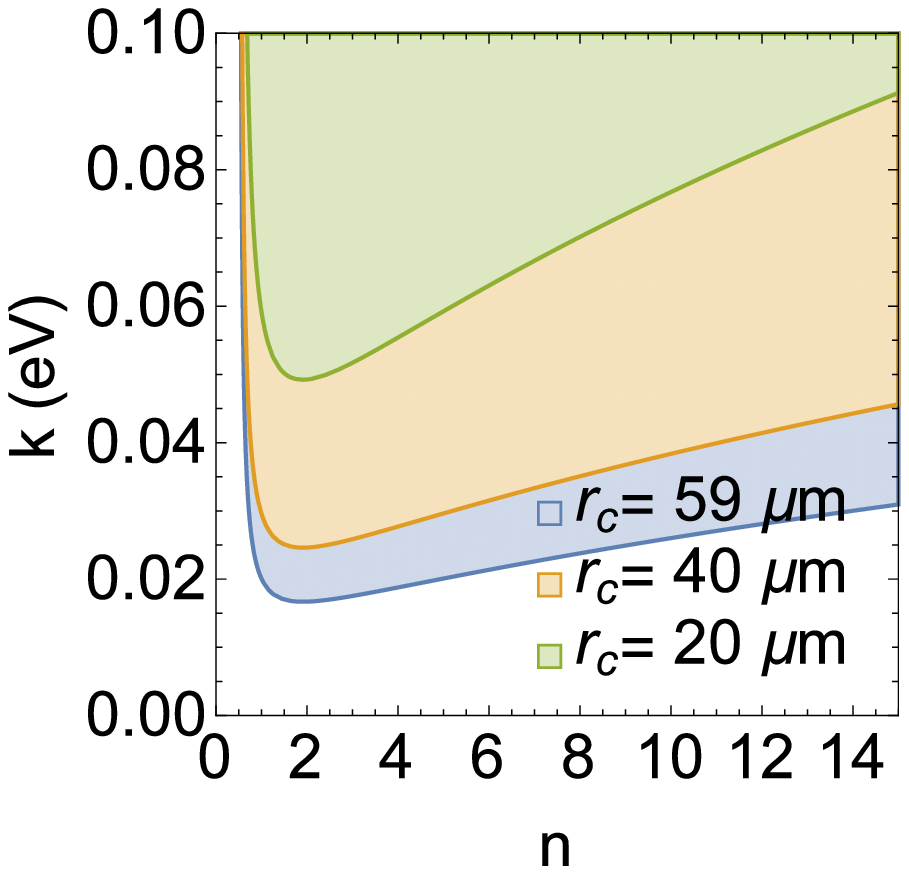}}
\subfigure[$L$]{\includegraphics[width=4cm]{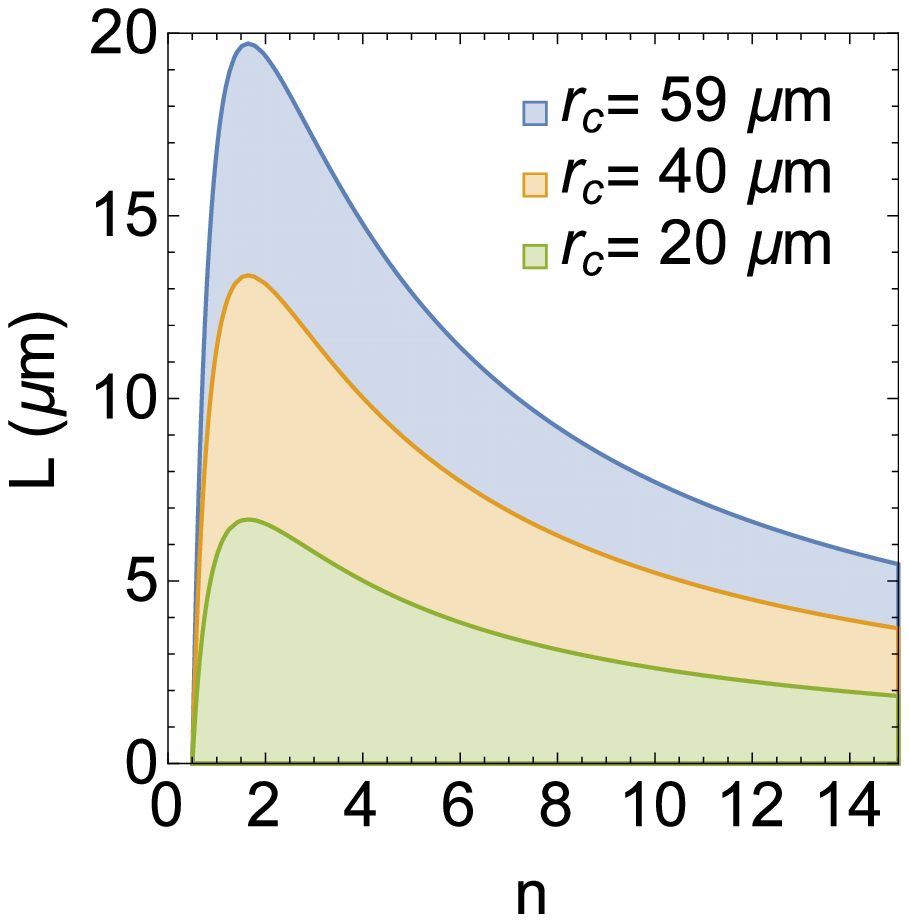}}
\subfigure[$m_{n'}$]{\includegraphics[width=4cm]{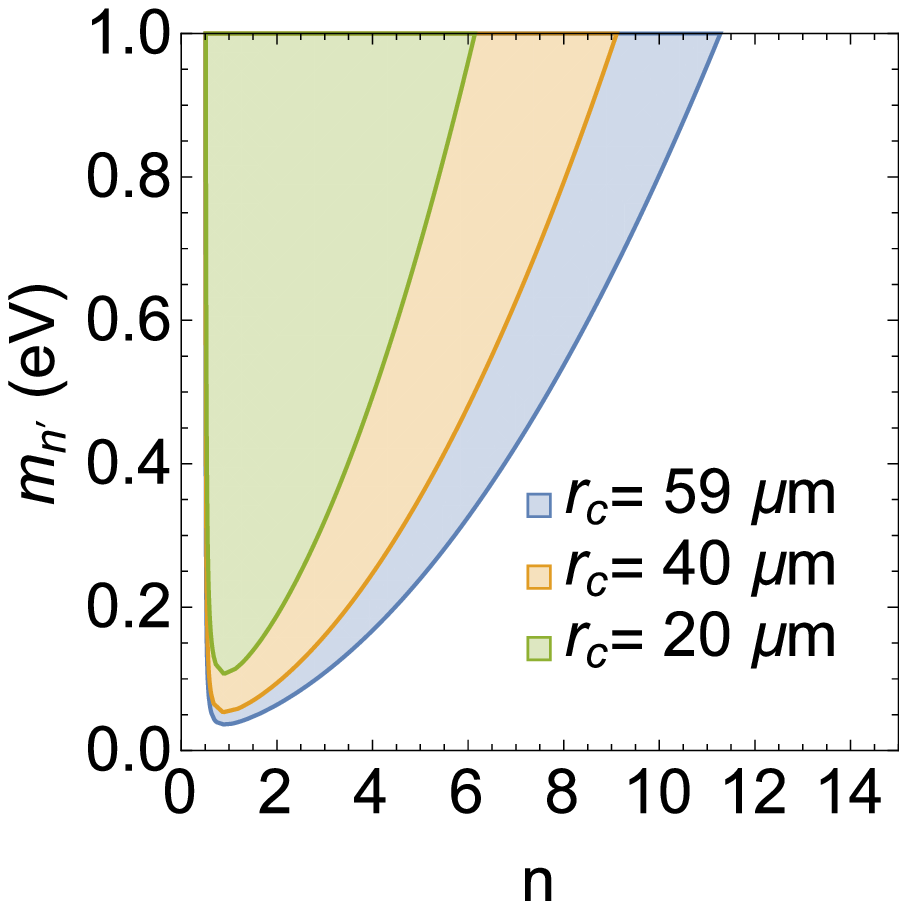}}
}
\caption{The constraints on (a) the parameter $k$, (b) the length of the extra dimension $L$, and (c) the mass of the $n'$th graviton KK mode $m_{n'}$. The areas with color show the allowed regions of $k$, $L$, and $m_{n'}$ with respect to different values of $r_{c}$.}\label{kLm}
\end{figure}

As is shown in Fig.~\ref{kLm}, the blue areas denote the allowed values of $k$, $L$, and $m_{n'}$, which are consistent with the recent test of the gravitational inverse-square law~\cite{Jun1}. The orange and green areas imply that, if the results of the future tests still obey the gravitational inverse-square law, the lower bounds on $k$ and $m_{n'}$ and the upper bound on $L$ are required to be larger and smaller, respectively. On the contrary, some parameters in the $f(|\hat{\Omega}|)$ brane could be constrained as long as the deviations from the gravitational inverse-square law are found on the future gravitational experiments. In addition, the deviations from the Standard Model found on the future experiments of high-energy particle collisions could also determine the mass spectrum of the KK gravitons, and therefore, together with Fig.~\ref{kLm}, constrain the free parameter $n$ in this brane model.

We finally conclude that, though the contributions from the massive gravitons have not been seen on the recent gravitational test, our model is still not excluded yet. From the theoretical side, this deviation is probably apparent with the separation at the length scale of a nanometer at most. Therefore, we still hopeful that the signal from the extra dimension can be obtained using the next generation of the gravitational test.

\section{Conclusion}\label{6}

In summary, in this paper, we investigated the thick $f(|\hat{\Omega}|)$ brane model and stability of the tensor perturbations of the brane system. We first reviewed EIBI theory and its generalized theory, i.e., $f(|\hat{\Omega}|)$ theory, in Palatini formalism. It is worth noting that the auxiliary tensor $p_{MN}=g_{MN}+bR_{MN}(\Gamma)$ is the auxiliary metric compatible with the independent connection $\Gamma$ if and only if $f(|\hat\Omega|)=|\hat\Omega|^{1/2}+\text{const.}$, where the constant term can be absorbed into the cosmological constant. In fact, this special choice corresponds to EIBI theory. In addition, we investigated the five-dimensional brane model in $f(|\hat{\Omega}|)$ theory and gave an analytic thick brane solution for $f(|\hat{\Omega}|)=|\hat{\Omega}|^{\frac{1}{2}+n}$. Although the scalar field generating the thick brane still has a kink configuration like in other gravity theories, the vacuum is located at $\phi=0$ but not at $\phi=\pm v_0$. It should be noted that our solution did require a constraint on the parameter $n$, i.e., $n>1/2$, to make itself  meaningful.
It is interesting that the energy density in this model goes to negative near the origin of the extra dimension, which is a little bit stranger than other gravity theories. Nevertheless, one can still judge the brane configuration from the localized energy density. We analyzed the physical curvature scalar of the bulk spacetime and concluded that the spacetime is asymptotically AdS.

We also investigated the tensor perturbations of the two metrics, and showed that the choice of TT gauge on the spacetime metric perturbations would lead to the TT gauge condition on the auxiliary metric perturbations. It is important to note that, unlike what we have assumed before the calculation of the linear perturbation equations, the tensor perturbations of the two metrics were found to be connected with each other and to be independent of $f(|\hat\Omega|)$. Also, from the obtained Schr\"{o}dinger-like equation of the extra-dimensional part of the graviton KK mode, we proved that no tachyon state exists and the tensor mode is stable. The massless graviton was found to be localized on the thick brane, so the four-dimensional Newtonian potential could be recovered. It is remarkable that a finite conformal extra-dimensional coordinate $z$ was found and all the KK gravitons were restricted to be bounded states. At last, we obtained the low-energy effective theory and the correction to the usual Newtonian gravitational potential on the brane and gave some constraints on the parameters in this generalized EIBI theory.

The localization of the fermion fields and the gravitons is a conspicuous problem in the braneworld model. Many efforts have been paid in this area, and it is now well known that the localization problem could be addressed well in some thick braneworld models~\cite{Fu1,Bazeia1,Jana1,Dubovsky1,Mouslopoulos1,Ringeval1,Bietenholz1,RandjbarDaemi1,Koley1,Melfo1,Tamvakis1,Almeida1,Liu1}. On the other hand, the investigation on the fermion resonances makes it possible to find the massive modes with finite lifetimes~\cite{Almeida1,Liu1,Farokhtabar1,Liu2,Zhao1,Xie1,Zhang1,KeYang4}. As we have mentioned, in principle, the constraints on the parameters could be obtained from the combination of the future particle physics experiments and gravitational experiments on the brane through the mass spectrum. Therefore, it is interesting to give further study to the localization mechanism and the resonance property of the fermion fields in this model.

\section*{Acknowledgements}

We are thankful to the referees' suggestions and criticisms. We also thank Bao-Min Gu, Wen-Di Guo, Hao Yu, and Zi-Qi Chen for helpful discussions. This work was supported by the National Natural Science Foundation of China (Grants No.~11522541, No.~11375075, and No. 11705070); the Fundamental Research Funds for the Central Universities (Grants No.~lzujbky-2018-k11 and No.~lzujbky-2017-it68); and the Strategic Priority Research Program on Space Science, the Chinese Academy of Sciences (Grant No. XDA15020701).

\end{document}